\documentstyle[aps]{revtex}

\newcommand{\be}{\begin{equation}}
\newcommand{\ee}{\end{equation}}
\newcommand{\bea}{\begin{eqnarray}}
\newcommand{\eea}{\end{eqnarray}}

\begin{document}
%\begin{flushright}
%hep-th/0306048
%\end{flushright}

\title{Gravitational Redshift in Einstein-Kalb-Ramond Spacetime and Randall-Sundrum
Scenario}
\author{Soumitra SenGupta\footnote{Electronic address: {\em
tpssg@iacs.res.in}}$^1$, and Saurabh Sur\footnote{Electronic
address: {\em saurabh@juphys.ernet.in}}$^2$} 
\address{{\rm $^1$}Department of Theoretical Physics, Indian Association for the 
Cultivation of Science,\\ Jadavpur, Kolkata 700 032, India} 
\address{{\rm $^{2}$}Department of Physics, Jadavpur University,
Kolkata 700 032,India} 
\maketitle 

\vspace{0.5cm}
\vspace{0.5cm}

\begin{abstract}
It is shown that the gravitational redshift as predicted by Einstein's theory, is modified in presence of second rank antisymmetric
tensor (Kalb-Ramond) field in a string inspired background spacetime.
In presence of extra dimensions, the Randall-Sundrum brane world scenario is found to play a crucial
role in suppressing this additional shift. The bound on the value of the  warp factor is determined from the
redshift data and is found to be in excellent agreement with that determined from the requirements of Standard model.

\end{abstract}
\bigskip

The massless antisymmetric tensor Kalb-Ramond (KR) field $B_{\mu\nu}$, that appears in 
the string spectrum \cite{string}, implements in a natural way the spacetime background 
for string theory possessing {\it torsion} in addition to curvature. Quantum consistency 
of the theory further demands the augmentation of the KR field strength $\partial_{[\mu} 
B_{\nu\lambda]}$ with the Chern-Simons (CS) three-form. The CS extension, in turn, plays 
the crucial role in restoring the $U(1)$ gauge symmetry \cite{pmssg} that is apparently 
lost when torsion is coupled minimally with the gauge field in the original Einstein-Cartan
theory. Solutions of the modified 
general relativistic field equations in various
situations may now provide new results leading to a possible quantitative assessment of 
string inspired torsion models. In an earlier work \cite{skssgss}, the study of geodesics 
in static spherisymmetric KR spacetimes with or without gravitating matter provided evidences 
for effects of the KR field in the context of bending of light trajectories as well as 
perihelion precession of planetary orbits which differ from the general relativistic estimates. 
Although the prospect of detecting the KR field is quite pronounced in an otherwise empty 
spacetime (which incidentally resembles a {\it wormhole} or a {\it naked singularity}
\cite{skssgss,ssgss}), in presence of gravitating matter the KR field produces little 
effects on the above-mentioned phenomena. One possible reason for such smallness of 
torsion (or, the KR field) can be given in the context of theories with large extra 
dimensions \cite{add,rs} where torsion is supposed to co-exist with gravity in the bulk, while 
all the standard model fields are confined to a 3-brane.  For models of Randall-Sundrum 
type \cite{rs}, it has been shown recently \cite{bmsenssg} that inspite of having the same 
status as gravity in the bulk, effects of massless torsion becomes heavily suppressed 
on the standard model brane, thus producing the illusion of a torsionless Universe. 

With an aim of estimating this arguably weak KR field we examine here it's effect on 
another important general relativistic phenomenon, namely the gravitational redshift 
of signals from distant radio sources. The spacetime structure is taken to be static 
spherically symmetric.The absence of any experimental evidence of the contribution to the redshift
due to the KR field as well as imperceptible ambient temperature of the KR field with respect to the Cosmic Microwave
Background temperature yield specific bounds on the KR field energy. From a 
higher dimensional point of view such bounds may enable one to estimate the compactification 
parameters in a certain scheme. We study such circumstances when torsion (or KR) field 
resides in the higher dimensional bulk spacetime alongwith gravity while all standard model 
fields are supposed to be confined to a 3-brane \cite{bmsenssg}.

According to the formalism in \cite{pmssg}, the four dimensional effective action for 
Einstein-Cartan-KR coupling with torsion (identified as the modified KR field strength) is given by
\be
S ~=~ \int~ d^{4}x ~\sqrt{-g}~\left\{\frac{R (g)}{\kappa^2} ~-~ \frac 1 {12} {\bf H}^2 \right\}
\ee
where $\kappa^2 = 8 \pi G$ is the gravitational coupling constant and the three-form ${\bf H}$ 
is the strength of the antisymmetric KR field $B_{\mu\nu}$ plus the $U(1)$ electromagnetic 
CS term:~ ${\bf H} = d {\bf B} + \kappa {\bf A \wedge F}$ (${\bf A}$ being the electromagnetic
field and ${\bf F} = d {\bf A}$ it's strength). The field equations 
are given by
\be
D_{\mu} H^{\mu \nu \lambda} ~=~ 0 ~;~~~~~
R_{\mu}^{~\nu} ~-~ \frac 1 2 \delta_{\mu}^{~\nu} R ~=~ \kappa^2 ~T_{\mu}^{~\nu~(KR)}
\ee
where $D$ stands for the covariant derivative defined in terms of the usual Christoffel 
connection and  $T_{\mu}^{~\nu~(KR)}$ , the KR energy-momentum tensor is given by
\be
T_{\mu}^{~\nu~(KR)} ~=~ - \frac 1 6 \left( \frac 1 2 \delta_{\mu}^{~\nu} H_{\alpha 
\beta \gamma} H^{\alpha \beta \gamma} ~-~ H_{\mu \alpha \beta} H^{\nu \alpha
\beta} \right) 
\ee

We consider a general static spherical symmetric metric structure
\be
ds^2 ~=~ B(r) dt^2 ~-~ \frac{dr^2}{A(r)} ~-~ r^2 (d \theta^2 + \sin^2 \theta d \phi^2).
\ee
Relating the KR field strength to the derivative of a pseudoscalar {\it axion} field 
$\xi$ via the duality: ~$\partial_{[\mu} B_{\nu \lambda]} = \epsilon_{\mu \nu 
\lambda}^{~~~~\sigma} \partial_{\sigma} \xi$, one can check that $\xi$
satisfies the massless Klien-Gordon equation ~$D_{\mu} D^{\mu} \xi = 0$.~ For the 
above metric this yields ~$(d\xi/dr)^2 = b/(4 \pi r^4 A B)$,~ where the integration 
constant $b$ determines the measure of the torsion energy density $\rho_{KR}$ which  is given by
\be
\rho_{KR} (r) ~=~ \frac{b}{4 \pi r^4 B(r)} ~+~ O(\kappa).
\ee
Dropping the $O(\kappa)$ contributions that arise by virtue of the Chern-Simons
extension, we write the gravitational field equations as
\bea
B (r A)' ~-~ B &=& -~\kappa^2 ~\frac{b}{8 \pi r^2} \\
A (r B)' ~-~ B &=& \kappa^2 ~\frac{b}{8 \pi r^2} \\
A B'' - \frac{A B'^2}{2 B} + \frac{A' B'} 2 + \frac{(A B)'} r &=& -~\kappa^2 ~\frac{b}
{4 \pi r^2}.
\eea
where a prime denotes differentiation with respect to $r$. 

Assuming general series' of the forms ~$B(r) = 1 + \sum_{i=1}^{\infty} b_i/r^i~;~
A(r) = 1 + \sum_{j=1}^{\infty} a_j/r^j$,~ and demanding asymptotic flatness
$B, A \rightarrow 1$ as $r \rightarrow \infty$, we obtain the general solution
for the metric coefficients $B$ and $A$ as
\bea
B(r) &=& 1 - \frac{2 m} r + b \left( \frac m {3 r^3}
+ \frac{2 m^2}{3 r^4} + \frac{24 m^3 - 3 m b}{20 r^5} + \cdots
\right) \\
A(r) &=& 1 - \frac{2 m} r + b \left( \frac 1 {r^2}
+ \frac m {r^3} + \frac{4 m^2}{3 r^4} + \frac{24 m^3 - m b}{12 r^5}
+\cdots \right). 
\eea
where $m$ is the usual gravitating mass and we have redefined $b$ by absorbing 
a factor of $G$ in it.

Now, for a typical atomic transition, the fractional shift $z$ in the frequency 
($\nu^{*}$) of a signal emitted on the surface 
of a distant object O$^*$ and the frequency ($\nu$) of the signal emitted via the 
same transition occurring at the observer's location O is given by the well-known 
formula \cite{wein}
\be
z ~\equiv~ \frac{\nu^{*} - \nu}{\nu} ~=~ \left\{ \frac{g_{tt} (r^{*})}{g_{tt} 
(r)} \right\}^{1/2} -~ 1
\ee
where $r^*$ is the radius of the distant object O$^*$ (which is considered to be spherical)
and $r$ is the distance between O$^*$ and O. 

In pure Schwarzschild geometry (without torsion) the frequency shift is given by
\be
z_s ~=~ \left( \frac m r - \frac m {r^*} \right) ~-~ \left( \frac{m^2}{r r^*} \right) ~+~
\left( \frac{3 m^2}{2 r^2} - \frac{m^2}{2 r^{*2}} \right) ~+~ \cdots.
\ee
Since $r^*$ is always less than $r$, the leading term on the right is negative implying
redshift.

In presence of torsion, an otherwise empty spacetime ($m = 0$) has the structure 
of a {\it naked singularity} or a {\it wormhole} (depending on the sign of $b$) 
\cite{skssgss,ssgss}, viz, $B = 1,~ A = 1 + b/r^2$, and the above formula yields zero 
frequency shift $z = 0$. However, in general, for non-zero $m$, 
considering the smallness of the torsion parameter $b$, we can relate the total 
frequency shift $z$ with the pure Schwarzschild shift $z_s$ by
\be
\left(\frac{1 + z}{1 + z_s}\right)^2 = \frac{1 + b f(r^*) + O(b^2)}
{1 + b f(r) + O(b^2)}~;~~~~~~~
f(r) = \frac 1 {3 r^2} \left( \frac m r + \frac{4 m^2}{r^2} + \cdots \right).
\ee
In most astrophysical observations, the source and destination objects O$^*$ and 
O are quite far apart, i.e., $r^* \ll r$. Also, for light to emit from O$^*$ its mass
$m < r^*$. Therefore we can approximately express the total frequency 
shift $z$ as
\be
z ~=~ z_s~\left(1 + \frac{\Delta z}{z_s}\right)~;~~~~~~ z_s \approx -\frac m {r^*} 
+ \cdots ~;~~~~~~
\frac{\Delta z}{z_s} \approx - \frac{b}{6 r^{*2}}
\ee
The maximum KR field energy density that can be obtained using the above
expression for the fractional departure $\Delta z/z_s$ is
\be
\mid \rho_{KR}^{max} \mid ~=~ \frac{b}{4 \pi G r^4} ~=~ 2~ \frac{M^*}{V^*}~ \left(
\frac{r^*} r \right)^4 \mid \Delta z/z_s^2 \mid
\ee
where $M^*$ and $V^*$ are the mass and volume of the source object O$^*$, $r^*$
is its radius and $r$ is the distance of the source O$^*$ from the observer O.

For the light coming from the sun to the earth, the general relativistic
predictions give $z_s \approx 2 \times 10^{-6}$ \cite{wein}. The maximum
KR field energy density that can be calculated using the standard solar system
data turns out to be
\be
\mid \rho_{KR}^{max} \mid ~ \sim ~ 7.5 ~\times~ 10^{16} ~\mid \Delta z/z_s \mid ~
J~ m^{-3}.
\ee
This is large enough unless the fractional change $\mid \Delta z/z_s \mid$
is sufficiently small and may lead to a very large alteration of the
energy distribution of the sun-earth system. One may as well expect
fair amount of modification due to the KR field on the redshift of
light from the luminous astrophysical sources giving rise to high local fluctuations on the
energy distribution in the Universe that may well exceed the highly
homogeneous Cosmic Microwave Background energy density ($\rho_{cmb}
~\sim~ 4 \times 10^{-14}~ J~ m^{-3}$ corresponding to the temperature $T_{cmb}
~\sim~ 2.7 K$). For light coming from the sun, lack of experimental evidences
in support of a huge background KR energy imposes an extremely low upper bound
on the fractional change on the redshift ($\mid \Delta z/z_s \mid$ less
than $\sim 10^{-32}$). This departure is well within the error bar for standard redshift
experiments \cite{will} ($e_z \sim 10^{-2}$) and its detection in present day experimental
setup is beyond question. 

In the context of the theories with large extra dimensions the above results are,
however, particularly useful especially in making predictions of the energy density of
the KR field that resides in the bulk alongwith gravity while the other standard model
fields are kept confined on a 3-brane. With reference to the analysis 
in \cite{bmsenssg} one finds that in a Randall-Sundrum (RS) 
scenario the zero mode of the KR field is suppressed enormously by the warp factor.
The standard
Kaluza-Klien (KK) decomposition of the bulk KR field is given by \cite{bmsenssg}
\be
B_{\mu\nu} (x,y) ~=~ \sum_{n=0}^{\infty} B_{\mu\nu}^n \frac{\chi^n (y)}{\sqrt{r_c}}
\ee
where $x$ stands for the usual 4-dim spacetime coordinates and $y$ is the extra dimension,
$r_c$ is the RS compactification radius and $\chi^n$ are the various KK modes. Using the
self-adjointness and normalization conditions the solution for the massless ($n = 0$) mode 
in 5-dim RS scenario is found to be
\be
\chi^0 ~=~ \sqrt{k r_c} ~e^{-k r_c \pi}
\ee
where $k$ is the higher dimensional Planck scale ($\sim 10^{18}~GeV$). In a 4-dim effective
theory the massless KR field thus gets suppressed by the factor of $\exp (-k r_c \pi)$ with 
respect to the 4-dimensionally projected massless gravitonic modes on the 3-brane. 
Accordingly, the
The duality relationship between the KR field strength and the axion $\xi~$ ($\partial_{[\mu}
B_{\nu\lambda]} = \epsilon_{\mu\nu\lambda}^{~~~~\sigma} \partial_{\sigma} \xi$) implies
the exponential suppression in the $\xi$-field as well. In a static spherically symmetric 
spacetime the maximum 4-dim effective KR field energy on the 3-brane is now suppressed from 
it's previous value $\rho_{KR}^{(0)~max}$ by   
\be
|\rho_{KR}^{max}| ~=~ e^{- 2 k r_c \pi} ~|\rho_{KR}^{(0)~max}|.
\ee
Estimates of $|\rho_{KR}^{(0)~max}|$ or, conversely, the warp factor $k r_c$ can 
now be made from above analysis as follows:

For the redshift of light coming from the sun, we consider the fractional correction 
$\mid \Delta z/z_s \mid$ due to the exponentially suppressed KR field on the 3-brane to be as 
small as $\sim 10^{-32}$ so that the maximum KR field energy $\mid \rho_{KR}^{max} \mid  \sim
10^{-14} J m^{-3} < \rho_{cmb}$, while the fractional correction $\mid \Delta z^{(0)}/z_s \mid$
without any Randall-Sundrum type suppression is of the order of the error bar
$\sim 10^{-2}$ \cite{will}. Plugging these values in
the above equation one finds $|\rho_{KR}^{(0)~max}| \sim 10^{15}~J~m^{-3}$ and $k r_c \approx
11$. This is in fairly good agreement with the value of the compactification radius needed in
5-dim RS scenario ($k r_c = 12$) to solve the hierarchy between Planck and electroweak scales. 
If, on the other hand, we take $k r_c = 12$ and $\mid \Delta z/z_s \mid \sim 10^{-32} ;~ 
\mid \Delta z^{(0)}/z_s \mid \sim 10^{-2}$, then the maximum KR field energy density reduces to
a fairly low value $\sim 10^{-12} J m^{-3}$ which falls well within the error bar of $\rho_{cmb}$.

We have thus shown that in a string inspired background spacetime the measure of gravitational redshift
departs significantly from that predicted by Einstein's theory because of the presence of
the second rank antisymmetric KR field in the background. The lack of any experimental support
in favour of such an additional shift can be explained by the presence of extra dimensions which in
a Randall-Sundrum brane world model suppresses this shift to a value smaller than the error bar of the
present redshift measurement experiments.The bound on the value of the warp factor for this suppression agrees remarkably well
with that already determined to resolve the well known naturalness problem in the Standard model.
We may therefore conclude that this excellent agreement of the values of the warp factor calculated in the
context of two totally different physical phenomena may be looked upon as an indirect support in favour of the
Randall-Sundrum brane world conjecture.
\vskip 0.2in 
\noindent
{\large{\bf Acknowledgment}}

SS acknowledges the Council of Scientific and
Industrial Research, Govt. of India for providing financial support.

\end{document}